\documentclass[conference]{IEEEtran}
\IEEEoverridecommandlockouts
\usepackage{cite}
\usepackage{amsmath,amssymb,amsfonts}
\usepackage{algorithmic}
\usepackage{graphicx}
\usepackage{textcomp}
\usepackage{xcolor}
\usepackage{bm}
\usepackage{booktabs}
\usepackage{array}
\usepackage{diagbox} 
\usepackage{caption}
\usepackage{colortbl}
\usepackage{tabu}
\usepackage{subfigure}
\usepackage[numbers,sort&compress]{natbib}
\def\BibTeX{{\rm B\kern-.05em{\sc i\kern-.025em b}\kern-.08em
		T\kern-.1667em\lower.7ex\hbox{E}\kern-.125emX}}{\tiny }
\begin{document}
	\title{{\huge \textcolor{black}{Carrier-Sense Multiple Access for Heterogeneous Wireless Networks Using Deep Reinforcement Learning}}\\
	}
	\author{\IEEEauthorblockN{Yiding Yu, Soung Chang Liew, Taotao Wang}
		\thanks{Y. Yu, S. C. Liew and T. Wang are with the Department of Information Engineering, The Chinese University of Hong Kong. Emails:\{yy016, soung, ttwang\}@ie.cuhk.edu.hk}
	}
	
	\maketitle
	
	\begin{abstract}
		This paper investigates a new class of carrier-sense multiple access (CSMA) protocols that employ deep reinforcement learning (DRL) techniques for heterogeneous wireless networking, referred to as carrier-sense deep-reinforcement learning multiple access (CS-DLMA). Existing CSMA protocols, such as the medium access control (MAC) of WiFi, are designed for a homogeneous network environment in which all nodes adopt the same protocol. Such protocols suffer from severe performance degradation in a heterogeneous environment where there are nodes adopting other  MAC protocols.  This paper shows that DRL techniques can be used to design efficient MAC protocols for heterogeneous networking. In particular, in a heterogeneous environment with nodes adopting different MAC protocols (e.g., CS-DLMA, TDMA, and ALOHA), a CS-DLMA node can learn to maximize the sum throughput of all nodes. Furthermore, compared with WiFi's CSMA, CS-DLMA can achieve both higher sum throughput and individual throughputs when co-existing with other MAC protocols. Last but not least, a salient feature of CS-DLMA is that it does not need to know the operating mechanisms of the co-existing MACs. Neither does it need to know the number of nodes using these other MACs. 
	\end{abstract}
	
	
	\section{Introduction}
	This paper investigates a new class of carrier-sense multiple access (CSMA) protocols based on deep reinforcement learning (DRL) for heterogeneous wireless networking, referred to as carrier-sense deep-reinforcement learning multiple access (CS-DLMA). We show that nodes adopting CS-DLMA can learn a medium access strategy that maximizes the sum throughput of a heterogeneous network consisting of nodes adopting different medium access control (MAC) protocols. Furthermore, CS-DLMA achieves this without prior knowledge of the participating MAC protocols in the heterogeneous network.
	
\begin{figure}[!t]
		\centering
		\includegraphics[scale=0.65]{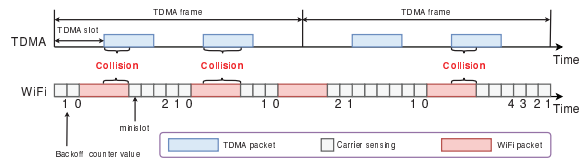}
		\caption{\textcolor{black}{Inharmonious co-existence of a TDMA node with a WiFi node. For simplicity, this example assumes each WiFi packet lasts four minislots, where a minislot is the slot used for carrier sensing by WiFi. This example also assumes each TDMA slot lasts four minislots. The TDMA node transmits packets in specific time slots within a TDMA frame repeatedly, from frame to frame, regardless of the MAC of the WiFi node. When
				the WiFi node senses the carrier to be idle and transmits in the subsequent four minislots, its transmission may collide with a TDMA packet that follows shortly, since the TDMA node does not perform carrier sensing before it transmits.}}
		\label{fig:illustration}
		\vspace{-0.22in}
	\end{figure}

	CSMA MAC protocols are widely used in practical networks today. However, these CSMA MACs are designed for homogeneous networks in which all nodes use the same CSMA MAC. A case in point is WiFi. The carrier sensing, collision avoidance, and binary exponential backoff mechanisms of WiFi \cite{bianchi2000performance} work well only if all nodes in the network adopt the same mechanisms. They do not work well in a heterogeneous network. To illustrate, consider the co-existence of a WiFi node and a node operating the time-division multiple access (TDMA) protocol. The TDMA node transmits in specific time slots in a frame consisting of multiple time slots, in a repetitive manner from frame to frame, as illustrated in Fig. 1. In particular, the TDMA channel access pattern is oblivious of the MAC protocol of WiFi; similarly, the MAC of WiFi is oblivious of the TDMA channel access pattern. As shown in Fig. \ref{fig:illustration}, the WiFi node may sense the channel to be idle and decide to transmit a packet, only to have the TDMA node transmit a packet shortly thereafter to result in a collision. A goal of CS-DLMA is to circumvent this problem through DRL. 
	
	In particular, we are interested in a ``model-free'' approach in which the CSMA protocol does not have detailed knowledge of the operating mechanisms of the other co-existing MAC protocols. Furthermore, the number of nodes operating each MAC protocol is also unknown. In other words, the CSMA node does not have a model that describes the heterogeneous environment precisely. The nodes operating CS-DLMA, referred to as DLMA nodes, must learn on the fly. Although there has been prior work on heterogeneous wireless networking, much prior work adopts a model-aware approach in which the knowledge of the co-existing MAC protocols is available – e.g., \cite{challita2018proactive} investigated an LTE network that employs DRL for harmonious co-existence with WiFi assuming full knowledge of the WiFi MAC mechanism. 
	
	In general, there are many DRL techniques \cite{li2017deep, sutton1998reinforcement}. In this paper, we focus on adapting deep Q-network (DQN) for use in CS-DLMA---DQN is a DRL technique originally proposed for Atari game playing in the seminal paper \cite{DQNpaper}. In \cite{DQNpaper}, one-step DQN was adopted, and in \cite{mnih2016asynchronous}, n-step DQN was adopted. In this paper, we study both one-step and n-step DQN (see Sections \ref{Sec:one-step DQN} and \ref{Sec:Methodology} for details). In addition, we propose and study a new variant of DQN, referred to as \textit{reward-backpropagation} DQN (RB-DQN). We show that RB-DQN can have better performance than one-step DQN and n-step DQN in a heterogeneous network consisting of ALOHA nodes, TDMA nodes, and DLMA nodes---specifically,  RB-DQN can achieve near-optimal sum throughput with faster convergence. 
	\subsection{Related Work}
	Since this paper focuses on MAC protocols that make use of DRL, we limit our review of related work in this domain only. The DRL MAC proposed in \cite{naparstek2017deep} is targeted for homogeneous wireless networks. Specifically, in \cite{naparstek2017deep}, multiple  nodes access multiple orthogonal channels using the same DRL MAC. By contrast, we focus on heterogeneous networks in which our CS-DLMA protocol must learn to co-exist with other MAC protocols. In this paper, we focus on the objective of maximizing the sum throughput of all nodes in the heterogeneous environment. The generalization of this objective will be explored in our future work. 
	
	The MAC in \cite{wang2018deep} and \cite{chang2018distributive} also concern multiple-channel access. Unlike \cite{naparstek2017deep}, the channels in \cite{wang2018deep} and \cite{chang2018distributive} are time-varying---each channel follows a two-state Markov chain. In particular, \cite{wang2018deep} assumes perfect spectrum sensing and multiple correlated channels, and \cite{chang2018distributive} assumes imperfect spectrum sensing and multiple orthogonal channels. In both \cite{wang2018deep} and \cite{chang2018distributive}, the nodes use DRL techniques to learn the system statistics (including state transition probabilities of the channels and spectrum sensing errors) to improve the spectrum utilization efficiency.  By contrast, our CS-DLMA uses DRL to learn the heterogeneous nodes' transmission patterns in time so that CS-DLMA nodes can schedule its own transmissions to achieve a certain system objective (\textcolor{black}{in this paper, we focus on the objective of maximizing the sum throughput}). 
	
	In \cite{challita2018proactive}, the authors investigated an LTE network that employs DRL for harmonious co-existence with WiFi. The focus of \cite{challita2018proactive} is to allow the downlinks of LTE base stations to use the WiFi channels in a non-disruptive way. Importantly, the scheme in \cite{challita2018proactive} is model-aware in that the LTE base stations know that the co-existing network is WiFi. By contrast, our CS-DLMA is model-free in that it does not presume knowledge of co-existing networks.
	
	In our previous work \cite{yu2018deep, yu2017deep}, we developed deep-reinforcement learning multiple access (DLMA) protocols for heterogeneous wireless networks. In \cite{yu2018deep, yu2017deep}, we assumed that nodes of different MACs use the same packet length. This assumption limits the application of DLMA in more general heterogeneous settings in which nodes of different MACs may adopt different packet lengths. This paper removes the same-packet-length assumption and introduces carrier sensing into DLMA. 
	
	\section{Reinforcement Learning Preliminaries}
	This section overviews the reinforcement learning techniques used in our CS-DLMA protocol. There are different techniques for reinforcement learning. This paper makes use of Q-learning \cite{sutton1998reinforcement, watkins1992q}. In the reinforcement learning (RL) framework, a decision-making agent interacts with an environment in discrete time steps \cite{sutton1998reinforcement}. At time step $ t $, the agent observes the environment state $ s_t $  and performs an action $ a_t $  chosen from an action set  $ A $ according to a policy  $ \pi $. The policy  $ \pi $ is a mapping from states to actions. Following action $ a_t $, the agent receives a reward $ r_{t+1} $  and the environment transits to state $ s_{t+1} $  in time step  $ t+1 $.
	
	\subsection{one-step Q-learning}
	Given a series of rewards, $r_{t + 1}, r_{t + 2}, \cdots $,  resulting from state-action pairs   $\left( s_t, a_t\right) , \left( s_{t+1}, a_{t+1}\right) , \cdots$, the accumulated discounted return going forward pinned at time  $ t $ is given by  $ {R_t} \buildrel \Delta \over = \sum\nolimits_{\tau  = t}^\infty  {{\gamma ^{\tau  - t}}{r_{\tau  + 1}}} $, where $\gamma  \in \left( {0,1} \right]$  is a discount factor. Because of the randomness in the state transitions, ${R_t}$   is in general  a random variable.  The expected accumulated discounted return of a state-action pair $ \left( s, a\right) $  of a policy  $ \pi $ is captured by a Q action-value function,    ${Q^\pi }\left( {s,a} \right) \buildrel \Delta \over = E\left[ {{R_t}|{s_t} = s,{a_t} = a,\pi } \right]$. The Q function associated with that of an optimal policy is   ${Q^*}\left( {s,a} \right) \buildrel \Delta \over = {\max _\pi }{Q^\pi }\left( {s,a} \right)$.  
	
	In Q-learning, the goal of the agent is  to learn the optimal policy in an online manner by observing the rewards while it takes action in successive time steps. In particular, the agent maintains the Q function,  $Q\left( s, a\right)$, for any state-action pair  $ \left( s, a\right) $,  in a tabular form. At time step  $ t $,  given state  $ s_t $, the agent selects an action  ${a_t} = \arg {\max _a}Q({s_t},a)$ based on its current Q table. This will cause the system to return a reward  $ r_{t+1} $  and move to state   $ s_{t+1} $. The experience at time step  $ t $  is captured by the quadruplet ${e_t} = ({s_t},{a_t},{r_{t + 1}},{s_{t + 1}})$. At the end of time step   $ t $, experience $ e_t $  is used to update entry $({s_t},{a_t})$ in $Q({s_t},{a_t})$  as follows: 
	\begin{align}\label{Qupdate}
	{Q_{new}}&\left( {{s_t},{a_t}} \right) \leftarrow {Q_{old}}\left( {{s_t},{a_t}} \right) +\nonumber \\
	&\beta \left[ {{r_{t + 1}} + \gamma \mathop {\max }\limits_{a'} {Q_{old}}\left( {{s_{t + 1}},a'} \right) - {Q_{old}}\left( {{s_t},{a_t}} \right)} \right],
	\end{align}
	
	The above is a smoothing operation that combines the old Q value,  $ {Q_{old}}\left( {{s_t},{a_t}} \right) $, with the new sample of expected return based on $ r_{t+1} $  and  $ s_{t+1} $,  $ {r_{t + 1}} + \gamma {\max _{a'}}{Q_{old}}\left( {{s_{t + 1}},a'} \right) $, to arrive at a new Q value. The parameter  $\beta  \in (0,1]$ captures the learning rate: in the computation of the new Q value, the relative weight of the old Q value is $1 - \beta $ and the relative weight of the new Q value sample is  $ \beta $.  
	
	As a variation to \eqref{Qupdate}, the  $ \varepsilon $-greedy algorithm is often adopted in action selection. Specifically, for the $ \varepsilon $-greedy algorithm, the action  ${a_t} = \arg {\max _a}Q({s_t},a)$  is  chosen with probability   $ 1-\varepsilon $; and a random action is chosen uniformly from the action set with probability $ \varepsilon $.  This is to avoid the algorithm from zooming in to a local optimal policy and to allow the agent to explore a wider spectrum of different actions in search of the optimal policy \cite{sutton1998reinforcement}. 
	
	Note that Q-learning is a \textit{model-free} learning framework in that it tries to learn the optimal policy without having a model that describes the operating behavior of the environment beyond what can be observed through the experiences. 
	\subsection{n-step Q-learning}
	The above Q-learning is referred to as one-step Q-learning because it updates  $Q\left( {{s_t},{a_t}} \right)$  based on the \textit{one-step return},  $ {r_{t + 1}} + \gamma {\max _{a'}}{Q_{old}}\left( {{s_{t + 1}},a'} \right) $ \cite{sutton1998reinforcement} . One drawback of one-step Q-learning is that the reward  ${r_{t + 1}}$ only directly affects  $Q\left( {{s_t},{a_t}} \right)$, and not  $Q\left( {{s_{t - 1}},{a_{t - 1}}} \right),Q\left( {{s_{t - 2}},{a_{t - 2}}} \right), \cdots $,  in previous time steps.  The values of other state-action pairs are affected only indirectly through the updated value of $Q\left( {{s_t},{a_t}} \right)$ in later learning steps. This can potentially slow down the learning process, since many updates are required to propagate a reward to relevant preceding states and actions. One way to speed up the propagation of rewards  is to use \textit{n-step  return} \cite{sutton1998reinforcement, peng1994incremental, mnih2016asynchronous}. In n-step Q-learning, we set  $Q\left( {{s_t},{a_t}} \right) = {r_{t + 1}} + \gamma {r_{t + 2}} + {\gamma ^2}{r_{t + 3}} +  \cdots  + {\gamma ^{n - 1}}{r_{t + n}} + {\gamma ^n}{\max _{a'}}Q\left( {{s_{t + n}},a'} \right)$. This results in a single reward ${r_{t + 1}}$  directly affecting the values of $ n $  preceding state action pairs. 
	\subsection{Deep Q-Network}\label{Sec:one-step DQN}
	It has been shown that in a stationary environment that can be fully captured by a Markov decision process, the Q values will converge to the optimal  ${Q^*}\left( {s,a} \right)$  if  the learning rate decays appropriately and each action in the state-action pair $ \left( s, a\right)  $ is executed an infinite number of times in the process \cite{sutton1998reinforcement, watkins1992q}. For many real-world problems, the state-action space for $ \left( s, a\right)  $ can be huge that the tabular update method, which updates only one entry in $Q\left( {s,a} \right)$  in each time step, can take an excessive amount of time for $Q\left( {s,a} \right)$ to converge to  ${Q^*}\left( {s,a} \right)$. If the environment changes in the meantime, convergence can never be attained. To allow fast convergence, function approximation methods are often used to approximate the Q values \cite{sutton1998reinforcement}.
	
	The seminal work \cite{DQNpaper} put forth deep Q-network (DQN), wherein a deep neural network model is used to approximate the action-value function  Q.  For simplicity, we refer to the neural network in DQN as QNN. The input to QNN is a state  $ s $, and the outputs are the approximated Q values for different actions,  $ \left\{ {Q\left( {s,a;{\bm{\theta }}} \right)|a \in A} \right\} $, where  ${\bm{\theta }}$ is a parameter vector consisting of the weights of the edges in the neural network.  For action execution, the  $ \varepsilon $-greedy algorithm based on the approximated Q values is adopted. For training, the parameter vector ${\bm{\theta }}$  is updated by minimizing the following loss function:
	\begin{align}\label{loss1}
	L\left( \bm{\theta } \right) = \frac{1}{N_E} \sum\limits_{e_i \in E} & \bigg[ \Big( r_{i + 1} + \gamma \mathop {\max }\limits_{a'} Q\left( s_{i + 1},a';\bm{\theta }^-  \right) -   \nonumber  \\
	& 	Q\left( s_i,a_i;\bm{\theta} \right) \Big)^2 \bigg].
	\end{align}
	
	There are two important  ingredients in DQN. The first ingredient is experience replay \cite{lin1992self, DQNpaper}. Instead of training QNN with a single experience associated with one action execution, multiple experiences could be pooled together for  batch training. In particular, an experience buffer  stores a fixed number of experiences gathered from different time steps. For a round of training, a minibatch $ E $  consisting of $ N_E $  random experiences taken from the experience buffer  is used in the computation of \eqref{loss1}. The second ingredient is the use of a separate ``target'' neural network in the computation of  $ {r_{i + 1}} + \gamma {\max _{a'}}Q\left( {{s_{i + 1}},a';{{\bm{\theta }}^ - }} \right) $, in \eqref{loss1}. In particular, the target neural network's parameter vector  is  ${{\bm{\theta }}^ - }$ rather than ${\bm{\theta }}$  in the QNN being trained. This separate target neural network is named target QNN and is a copy of a previously used QNN: the parameter ${{\bm{\theta }}^ - }$  of target QNN is updated to the latest ${\bm{\theta }}$ of QNN once in a while. 
	
	We refer to the above DQN algorithm as one-step DQN. The extension to the n-step DQN algorithm is obvious (details to be given in Section \ref{Sec:Methodology}). 
	
	\section{CS-DLMA}\label{Sec:CS-DLMA}
	This section specifies the system model and the methodology of CS-DLMA investigated in this paper. 
	\subsection{System Model}\label{SystemModel}
	We consider time-slotted heterogeneous wireless networks in which different nodes transmit packets to an access point (AP) via a shared wireless channel. In this paper, we consider four types of networks whose nodes use different protocols: (i) CS-DLMA, (ii) WiFi (more exactly, a simplified WiFi-like CSMA protocol), (iii) TDMA, and (iv) different variants of ALOHA. Among them, CS-DLMA and WiFi nodes have the capability for carrier sensing, while TDMA and ALOHA nodes do not. 
	
	We assume different networks may have different slot granularities. The smallest slot is the basic slot used by DLMA nodes to perform carrier sensing or to transmit packets. The basic slot is also used by WiFi nodes to perform carrier sensing. WiFi slot, TDMA slot and ALOHA slot consist of multiple basic slots and are used by WiFi nodes, TDMA nodes and ALOHA nodes to transmit packets, respectively (i.e., a WiFi/TDMA/ALOHA packet lasts a duration of a WiFi/TDMA/ALOHA slot). We denote the ratio of WiFi slot, TDMA slot and ALOHA slot to the basic slot by  $ R_W $, $ R_T $ and  $ R_A $. We assume a node can  begin transmission only at the beginning of its own packet slot and must finish the transmission at the end of this packet slot. Simultaneous transmissions by multiple nodes result in a collision. A packet transmitted without collision is successfully received by the AP. After each successful transmission, the AP  broadcasts an acknowledgment that contains the packet length information, interpreted as a ``reward'' in RL, as will be elaborated later in this subsection. 
	\begin{table*}
		\renewcommand\arraystretch{1.4}
		\caption{MAC mechanisms of different nodes.}
		\label{RadioNodes}
		\vspace{-0.1in}
		\flushleft
		\scalebox{1}{
			\begin{tabular}{ |p{1.7cm}|p{15.5cm} |}
				\hline
				\textbf{Node Type} &\textbf{ Description} \\ \hline \hline
				{\footnotesize \textbf{TDMA}} &  \textit{{\footnotesize A TDMA node transmits in $ X $ specific TDMA slots within a TDMA frame of $ Y $  TDMA slots in a repetitive manner from frame to frame.}} \\ \hline 
				{\footnotesize \textbf{$ q $-ALOHA}}  & \textit{{\footnotesize A  $ q $-ALOHA node transmits with a fixed probability  $ q $ in each ALOHA slot in an i.i.d. manner from ALOHA slot to ALOHA slot.}} \\ \hline
				{\footnotesize \textbf{Fixed-window ALOHA}} & \textit{{\footnotesize A fixed-window ALOHA (FW-ALOHA) node generates a random counter value  $w \in \left[ {0,W - 1} \right]$ after it transmits in an ALOHA slot. It then waits for  $w$ ALOHA slots before its next transmission. The parameter $ W $  is referred to as the window size.}}   \\
				\hline
				{\footnotesize \textbf{Exponential-backoff ALOHA}}& \textit{{\footnotesize  Exponential backoff ALOHA (EB-ALOHA) is a variant of FW-ALOHA that uses a \textit{binary exponential backoff} mechanism, in which the window size is doubled each time its transmission incurs a collision up to a maximum window size of  ${2^m}W$, where $ m $  is the ``maximum backoff stage''. Upon a successful transmission, the window size reverts to the initial window size  $ W $.}} \\
				\hline
				\hline
				{\footnotesize \textbf{WiFi}} & \textit{{\footnotesize A WiFi node is a node that employs a CSMA/CA protocol with the same backoff counter and binary exponential backoff mechanisms as EB-ALOHA. Before transmitting a packet, the WiFi node performs carrier sensing on a per basic slot basis. For each basic slot the channel is sensed idle, the backoff counter is decreased by one. The countdown of the counter is frozen if the channel is sensed busy. When the counter value reaches zero, the WiFi node transmits.}}\\
				\hline
				{\footnotesize \textbf{DLMA}} & \textit{{\footnotesize A DLMA node is a node that uses our CS-DLMA protocol to decide whether to transmit in a basic slot. If it transmits, it will get an immediate feedback from the AP indicating whether the transmission is successful or not; if it does not, it will perform carrier sensing to check if the channel is busy or idle.}}    \\
				\hline
		\end{tabular}}
		\vspace{-0.2in}
	\end{table*}
	
	\textcolor{black}{Table \ref{RadioNodes} summarizes the MAC mechanisms of different nodes. Part of  this paper will investigate and compare ``co-existence of DLMA with TDMA and ALOHA'' with  ``co-existence of WiFi with TDMA and ALOHA'' (see Section \ref{Sec:eva2}). }

	We now give the details of CS-DLMA. To transform the medium access problem faced by a DLMA node to a reinforcement learning problem, we need to define the corresponding \textit{action}, \textit{state}, and \textit{reward} in RL. 
	
	The \textit{action} taken by a DLMA node in basic slot  $ t $ is   $ a_t \in$ \{\textit{TRANSMIT}, \textit{SENSE}\}, where \textit{TRANSMIT} means that the DLMA node transmits, and \textit{SENSE} means that it performs carrier sensing (i.e., it does not transmit). If  $ a_t =$ \textit{TRANSMIT}, the agent will get an observation  \textit{SUCCESSFUL} or \textit{COLLIDED} from the AP, indicating whether the packet is successfully transmitted or not; if   $ a_t =$ \textit{SENSE}, the agent will get an observation  $ z_t =$  \textit{BUSY} or \textit{IDLE}, indicating whether the channel is being occupied or not occupied by other nodes. We define the \textit{channel state} in basic slot $ t + 1 $  as the action-observation pair  ${c_{t + 1}} \buildrel \Delta \over = \left( {{a_t},{z_t}} \right)$. There are four possibilities for  $ c_{t+1} $: \{\textit{TRANSMIT}, \textit{SUCCESSFUL}\}, \{\textit{TRANSMIT}, \textit{COLLIDED}\}, \{\textit{SENSE}, \textit{BUSY}\} and \{\textit{SENSE}, \textit{IDLE}\}. We define the \textit{environmental state} in basic slot $ t+1 $  to be  ${s_{t + 1}} \buildrel \Delta \over = \left[ {{c_{t - M + 2}}, \cdots ,{c_t},{c_{t + 1}}} \right]$, where the parameter $ M $  is the state history length (number of past basic slots) to be tracked by the DLMA node. 
	
	After taking action  $ a_t $, a \textit{reward}  $ r_{t+1} $ is generated at the end of basic slot  $ t $ and the state becomes $ s_{t+1} $  in basic slot  $ t+1 $. If the channel is idle or there is a collision in basic slot  $ t $, then   $ r_{t+1}=0 $. For a successful transmission, the reward varies according to the length of the packet transmitted. In particular, if a DLMA node transmits a packet of one basic slot in duration, then     $ r_{t+1}=1 $; if a WiFi/TDMA/ALOHA node successfully completes the transmission of a packet lasting a few basic slots, then   $ r_{t+1}=R_W/R_T/R_A $ (e.g.,  for TDMA, if a TDMA packet begins transmission  in basic slot $ t-\left( R_T - 1\right)  $  and the transmission is completed successfully in basic slot  $ t $,  then the reward at the end of basic slot $ t $ is $ r_{t+1} = R_T $). Note that in this basic scheme, for a packet lasting more than one basic slot, the reward is given only at the end of the last basic slot of the successfully transmission, and no reward is given in the earlier basic slots. In this study, both one-step DQN and n-step DQN use this basic reward scheme. We will also introduce  and investigate another reward scheme called \textit{reward-backpropagation }that amortizes the reward over each and every basic slots during which the packet is in transmission. 
	\subsection{Methodology}\label{Sec:Methodology}
	In \cite{yu2018deep, yu2017deep}, we put forth DLMA protocols without carrier sensing for co-existence with different nodes transmitting packets of the same length. DLMA is based on one-step DQN in which the QNN is feedforward neural networks (FNN). 
	
	However, in our new setting here with introduction of carrier sensing and different slot lengths, we find that CS-DLMA using ``FNN + one-step DQN'' fails to learn an optimal strategy, as will be detailed in Section \ref{Sec:eva1}. As a potential solution, we put forth a \textit{reward-backpropagation} DQN (RB-DQN) algorithm that outperforms the original one-step DQN and n-step DQN. Furthermore, we explore the use of recurrent neural networks (RNN) as a replacement for FNN.  

\begin{figure}[!t]
	\centering
	\includegraphics[scale=0.55]{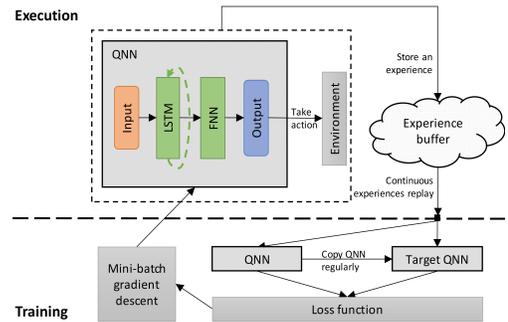}
	\caption[]{\textcolor{black}{Architecture of components realizing CS-DLMA.}\footnotemark}
	\label{fig:architecture}
	\vspace{-0.2in}
	\end{figure}

	\textcolor{black}{Fig. \ref{fig:architecture} shows the overall implementation architecture that realizes CS-DLMA assuming the QNN is an RNN. We now describe four key components in the architecture: (i) neural network, (ii) experience buffer, (iii) continuous experience replay and (iv) loss function. }
	\subsubsection{Neural Network}
	The RNN consists of an input layer, two hidden layers, and an output layer. The input to the RNN is the current state. The two hidden layers consist of  a long-short-term-memory (LSTM) \cite{hochreiter1997long} layer and an FNN layer. The outputs are the approximated Q values for different actions given the input state. 

	\footnotetext{For convenience, in our simulation, we assume execution of decisions and training of QNN are synchronous. In particular, the training is done at the end of each time step after an execution. In practice, execution and training can be done asynchronously and in parallel. A detailed discussion can be found in our paper \cite{yu2017deep}.} 
	
	In particular, the input to the input layer in basic slot $ t+1 $   is state  ${s_{t + 1}} = \left[ {{c_{t - M + 2}}, \cdots ,{c_t},{c_{t + 1}}} \right]$, where $ {c_{t + 1}} = \left( {{a_t},{z_t}} \right) $ is the channel state, and $ z_t $  in  $ c_{t+1} $ is the observation of the DLMA node with four possibilities: \textit{SUCCESSFUL}, \textit{COLLIDED}, \textit{BUSY} or \textit{IDLE}.  We adopt \textit{one-hot encoding} \cite{lecun2015deep} to encode these four possibilities. 
	
	Fig. 3 shows the difference between FNN-based QNN and RNN-based  QNN in processing   $ s_{t+1} $ received from the input layer. After receiving   $ s_{t+1} $, FNN processes it directly; by contrast, after receiving  $ s_{t+1} $, RNN processes the elements,  ${c_{t - M + 2}}, \cdots ,{c_t},{c_{t + 1}}$ in   $ s_{t+1} $ sequentially, keeping an internal state as it moves from one element to the next. In this way, the causal relationship between elements in  $ s_{t+1} $  (e.g., $ c_t $  precedes  $ c_{t+1} $) is explicitly embedded into the way RNN processes the input \cite{lecun2015deep}.  On the other hand, the causal relationship between elements in  $ s_{t+1} $ is not explicitly given to FNN. FNN will need to learn this relationship, if it manages to learn at all. 
	\begin{figure}[!t]
		\centering
	\subfigure[{\scriptsize FNN}]{
		\label{fig:FNN}
		\begin{minipage}[t]{0.1\textwidth}
			\centering
			\includegraphics[scale=0.5]{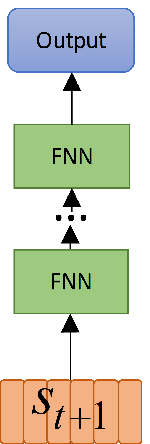}
			\vspace{-0.05in}
	\end{minipage}}
	\hspace{-2cm}
	\subfigure[{\scriptsize RNN}]{
		\label{fig:RNN}
		\begin{minipage}[t]{0.45\textwidth}
			\centering
			\includegraphics[scale=0.5]{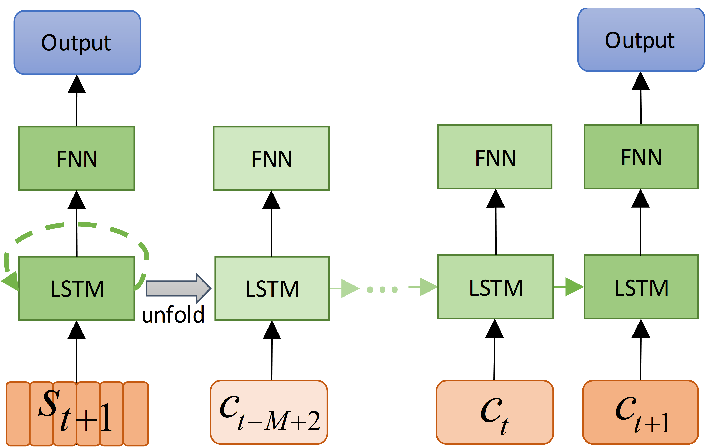}
			\vspace{-0.05in}
	\end{minipage}}
     \hspace{-2cm}
	\caption{{\scriptsize FNN-based QNN versus RNN-based QNN.} }
	\label{fig:FNN_RNN}
	\vspace{-0.2in}
	\end{figure}
	\subsubsection{Experience Buffer}
	In one-step DQN \cite{DQNpaper}, an experience is defined by the quadruplet $({s_t},{a_t},{r_{t + 1}},{s_{t + 1}})$  and is stored in the experience buffer after each interaction between the agent and the environment. In n-step DQN and RB-DQN, there are some modifications. 
	\begin{itemize}
		\item \underline{n-step DQN.} The experience is redefined as $({s_t},{a_t},{r_{t + 1}},{r_{t + 2}}, \ldots ,{r_{t + n}},{s_{t + n}})$  in order to compute the n-step return in the loss function (given in the later part of this subsection). 
		\item \underline{RB-DQN.} After storing  $({s_t},{a_t},{r_{t + 1}},{s_{t + 1}})$ into the experience buffer, a \textit{reward-backpropagation} mechanism is performed. This mechanism first checks  the value of  ${r_{t + 1}}$. If  ${r_{t + 1}} = R\left( {R > 1} \right)$,  then it sets   ${r_{t + 1}} = {r_t} =  \cdots  = {r_{t - R + 2}} = 1$, amortizing and backpropagating the reward to experiences of earlier time steps. If $ r_{t+1} = 1 $ or $ 0 $, then do nothing. 
	\end{itemize}
	
	For implementation, it is inefficient to store an experience in the form of $({s_t},{a_t},{r_{t + 1}},{s_{t + 1}})$  or $({s_t},{a_t},{r_{t + 1}},{r_{t + 2}}, \ldots ,{r_{t + n}},{s_{t + n}})$  since two consecutive experiences have many common elements. For example,  $ s_{t+1} $ in $({s_{t + 1}},{a_{t + 1}},{r_{t + 2}},{s_{t + 2}})$  is only a time-shift version of $ s_t $  in  $({s_t},{a_t},{r_{t + 1}},{s_{t + 1}})$. It is superfluous to store the overlapped elements for both experiences. A more efficient implementation is to store the abbreviated experience  $\left( {{c_t},{a_t},{r_{t + 1}},{c_{t + 1}}} \right)$. The complete experience $({s_t},{a_t},{r_{t + 1}},{s_{t + 1}})$  or  $({s_t},{a_t},{r_{t + 1}},{r_{t + 2}}, \ldots ,{r_{t + n}},{s_{t + n}})$ can be obtained from consecutive abbreviated experiences by means of \textit{continuous experience replay} (detailed in the next paragraph). Note that for n-step DQN, we do not need to redefine an abbreviated experience  $\left( {{c_t},{a_t},{r_{t + 1}},{c_{t + 1}}} \right)$ to  $\left( {{c_t},{a_t},{r_{t + 1}},{r_{t + 2}}, \ldots ,{r_{t + n}},{c_{t + n}}} \right)$. For RB-DQN, the \textit{reward-backpropagation} mechanism is still necessary.
	\subsubsection{Continuous Experience Replay}
\begin{figure}[!t]
		\subfigure[{\scriptsize one-step DQN or RB-DQN}]{
			\label{fig:sample1}
			\begin{minipage}[t]{0.23\textwidth}
				\centering
				\includegraphics[scale=0.32]{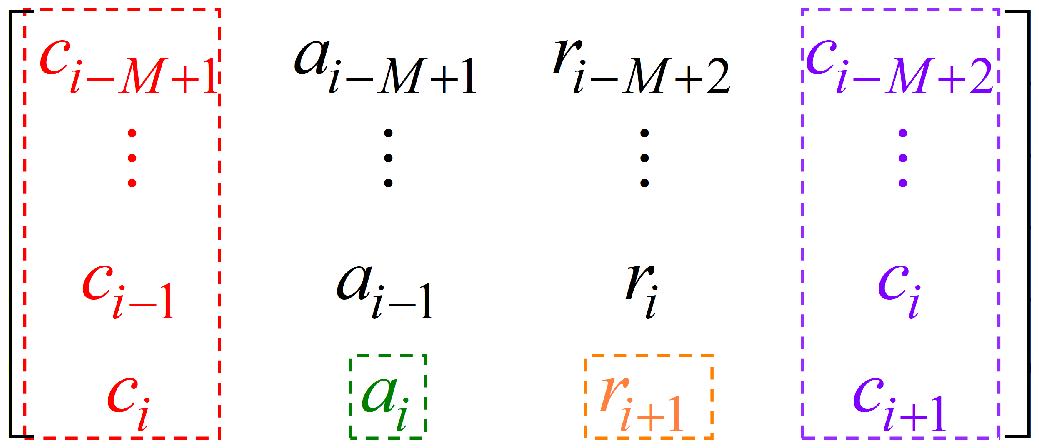}
				\vspace{-0.05in}
		\end{minipage}}
		\hspace{-0.2cm}
		\subfigure[{\scriptsize n-step DQN}]{
			\label{fig:sample2}
			\begin{minipage}[t]{0.24\textwidth}
				\centering
				\includegraphics[scale=0.37]{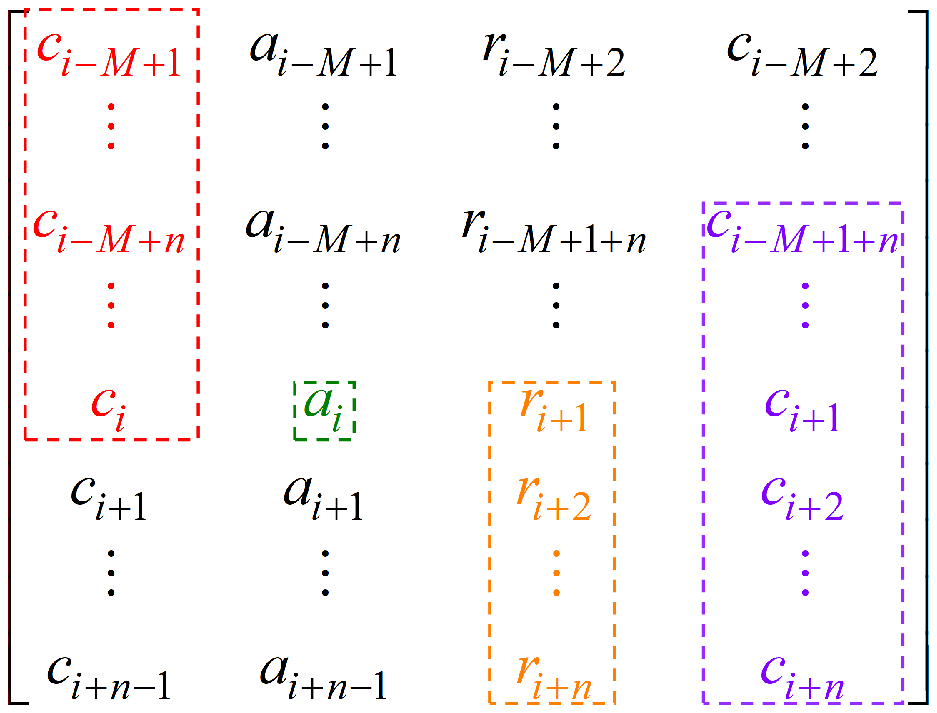}
				\vspace{-0.05in}
		\end{minipage}}
		\caption{{\scriptsize A sample in continuous experience replay.} }
		\label{fig:sample}
		\vspace{-0.2in}
	\end{figure}

	In conventional experience replay \cite{lin1992self, DQNpaper}, random experiences are sampled from the experience buffer to compute the loss function, with each sample being an experience  $({s_t},{a_t},{r_{t + 1}},{s_{t + 1}})$. After downsizing the experience to   $\left( {{c_t},{a_t},{r_{t + 1}},{c_{t + 1}}} \right)$, we will sample continuous experiences instead to extract the information necessary for computing the loss function. For one-step DQN and RB-DQN, as illustrated in Fig. \ref{fig:sample1}, each sample contains  $ M $ continuous experiences, and we extract $ {s_i} = \left[ {{c_{i - M + 1}}, \cdots ,{c_i}} \right] $,  $ a_i $,  $ r_{i+1} $,  $ {s_{i + 1}} = \left[ {{c_{i - M + 2}}, \cdots ,{c_{i + 1}}} \right] $ from it. For n-step DQN, as illustrated in Fig. \ref{fig:sample2}, each sample contains  $M + n - 1$ continuous experiences, we extract  $ {s_i} = \left[ {{c_{i - M + 1}}, \cdots ,{c_i}} \right] $,  $ a_i $,  $ {r_{i + 1}},{r_{i + 2}}, \ldots ,{r_{i + n}} $,  $ {s_{i + n}} = \left[ {{c_{i - M + 1 + n}}, \cdots ,{c_{i + n}}} \right] $ from it.
	\subsubsection{Loss Function}
	The loss function \eqref{loss1} is only suitable for one-step DQN and RB-DQN. A more general loss function that takes the n-step return into consideration is given by:
	\begin{align}\label{loss2}
	L\left( \bm{\theta } \right)  = \frac{1}{N_E} \sum\limits_{{e_i} \in E} & \bigg[ \Big( \sum\limits_{k = 0}^{n - 1} {{\gamma ^k}{r_{i + k + 1}}}  + \gamma^n\mathop {\max }\limits_{a'} Q\left( s_{i + n},a';\bm{\theta}^ - \right) \nonumber  \\
	& 	- Q\left( s_i,a_i;\bm{\theta } \right) \Big)^2 \bigg]
	\end{align}
	where ${e_i} = \left( {{s_i},{a_i},{r_{i + 1}}, \ldots ,{r_{i + n}},{s_{i + n}}} \right)$. When  $ n=1 $, \eqref{loss2} is the same as \eqref{loss1}.  
	With a loss function definition \eqref{loss1} or \eqref{loss2}, a minibatch gradient descent algorithm \cite{lecun2015deep} can then be used to update parameter $\bm{\theta}$. 
	
	\section{Performance Evaluation}
	This section evaluates the performance of CS-DLMA. First, we describe the simulation setup, including the values of the hyperparameters used in DQN, the performance metric, and the benchmark.  Second, we compare the performances of variants of CS-DLMA with different neural networks and different DQN implementations. Third, we compare the performances of CS-DLMA and WiFi when they co-exist with ALOHA and TDMA. Finally, we present detailed performance results of ``RNN + RB-DQN'', the best-performing  CS-DLMA variant studied in this paper, under different heterogeneous network settings.
	\subsection{Simulation Setup}
	\subsubsection{Hyperparameters} 
	As shown in Fig. \ref{fig:FNN_RNN}, the RNN has two hidden layers: one LSTM layer followed by one FNN layer. The number of neurons for each layer is 64 and the activation functions are \textit{ReLU} \cite{lecun2015deep}. We use \textit{RMSProp} \cite{tieleman2012lecture} to conduct minibatch gradient descent on \eqref{loss1} or \eqref{loss2}. \textcolor{black}{Since we assume CS-DLMA does not know the mechanisms of the co-existing MACs, we  use a relatively  large $M$ to cover a longer history so as to learn the behavior of potentially complex MACs (although in actuality, the MACs that we study here are not that complex and a small $M$ may suffice).  Specifically, for our simulations, we set $M = 40$.} To prevent the algorithms from getting stuck with a suboptimal decision policy before they gather enough  experiences, we apply an exponential decay  $ \varepsilon $-greedy method:  $ \varepsilon $ is initially set to 0.1 and $ \varepsilon $  decays by a multiplicative factor of 0.995 every basic slot until its value reaches 0.005. The values of hyperparameters are summarized in Table \ref{tab:hyperparameters}. 
	\begin{table}[htbp]
		\centering
		\caption{CS-DLMA Hyperparameters}
		\label{tab:hyperparameters}	
		\begin{tabular}{ccc}
			\toprule
			Hyperparameter& Value\\
			\midrule
			State history length  $ M $& 40\\
			$ \varepsilon $ in $ \varepsilon $-greedy algorithm & 0.1 to 0.005\\
			Discount factor $ \gamma $ & 0.9\\
			Experience buffer size & 500\\
			Experience-replay minibatch size $ N_E $ & 32\\
			Target network update frequency  & 200\\
			\bottomrule
		\end{tabular}
		\vspace{-0.05in}
	\end{table}
	\begin{figure*}[htbp]
		\subfigure[{\scriptsize \textcolor{black}{Different DQNs using two-hidden-layer FNN}}]{
			\label{fig:eva1_1_a}
			\begin{minipage}[t]{0.33\textwidth}
				\centering
				\includegraphics[scale=0.33]{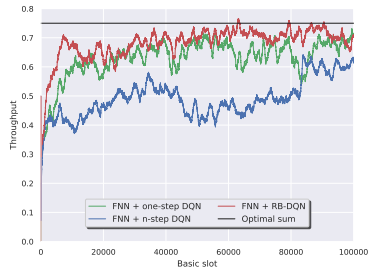}
				\vspace{-0.4in}
		\end{minipage}}
		\hspace{-0.3cm}
		\subfigure[{\scriptsize \textcolor{black}{Different DQNs using RNN}}]{
			\label{fig:eva1_1_b}
			\begin{minipage}[t]{0.33\textwidth}
				\centering
				\includegraphics[scale=0.33]{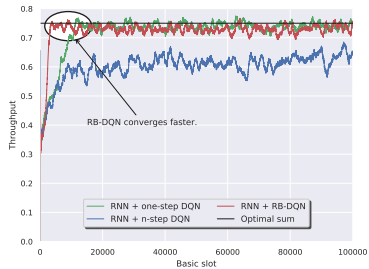}
				\vspace{-0.4in}
		\end{minipage}}
		\hspace{-0.3cm}
		\subfigure[{\scriptsize \textcolor{black}{RB-DQN using RNN and RB-DQN using FNNs of different numbers of hidden layers}}]{
			\label{fig:eva1_2}
			\begin{minipage}[t]{0.33\textwidth}
				\centering
				\includegraphics[scale=0.33]{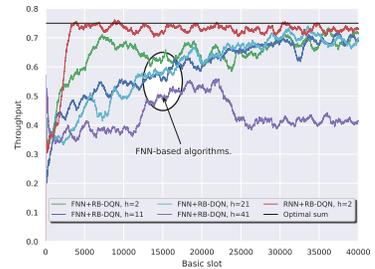}
				\vspace{-0.4in}
		\end{minipage}}
		\caption{\textcolor{black}{Short-term sum throughputs when one DLMA node (using different CS-DLMA algorithms) co-exists with one $ q $-ALOHA node and one TDMA node. Each line (except the black line) is averaged over four different runs.}}
		\label{fig:eva1}
		\vspace{-0.15in}
	\end{figure*}
	\subsubsection{Performance Metrics}
	In this paper, the objective of the DLMA node is to maximize the overall sum throughput. The throughput is defined by $\sum\nolimits_{\tau  = t - N + 1}^t {{r_\tau }} /N$, where  $ N $ is the smoothing window size. In our performance study, for  ``\textbf{short-term throughput}'' at basic slot $ t $,  $ N $ is set to 1000 (i.e., the throughput averaged over the past 1000 basic slots); for ``\textbf{long-term cumulative throughput}'' at basic slot $ t $, $ N $  is set to  $ t $  (i.e., the throughput averaged from basic slot 0 to basic slot $ t - 1$). 
	\subsubsection{Benchmark}
	The benchmark used in this paper is the optimal sum throughput that can be achieved by a model-aware node. The model-aware node knows the MAC mechanisms of co-existing nodes as well as the number of nodes executing each MAC protocol. For example, for co-existence with TDMA and ALOHA, the model-aware node knows the time slots during which TDMA nodes transmit and the random-access mechanism of the ALOHA nodes, as well as the number of TDMA nodes and the number of ALOHA nodes. The model-aware node executes an optimal MAC that maximizes the sum throughput based on this knowledge. The derivations of the optimal MAC and the associated sum throughput are given in \cite{benchmark}; we omit the derivations here to save space.
	\subsection{Different Variants of CS-DLMA}\label{Sec:eva1}
	We first present performance results of different variants of CS-DLMA under a specific heterogeneous network setting. In particular, we consider the co-existence of one DLMA node with one $ q $-ALOHA node and one TDMA node.
	
	
	The transmission probability $ q $  of the $ q $-ALOHA node is 0.4; the TDMA node occupies 2 TDMA slots within a TDMA frame of 5 TDMA slots. Both $ q $-ALOHA and TDMA have a packet length of 4 basic slots in this study, i.e., ${R_A} = {R_T} = 4$ (we will study $ q $-ALOHA and TDMA with different packet lengths later). For n-step DQN, we set  $ n=4 $, i.e., $ n $  equals to the packet length of $ q $-ALOHA and TDMA nodes. 
	
	Fig. \ref{fig:eva1} presents the short-term sum throughputs of different variants of CS-DLMA algorithms studied here. We also present the optimal sum throughput achieved by a model-aware node when it replaces the DLMA node. For the results in Fig. \ref{fig:eva1_1_a} and Fig. \ref{fig:eva1_1_b}, the numbers of hidden layers in both FNN and RNN are 2---the only difference is that the first hidden layer of RNN is LSTM (see Fig. \ref{fig:FNN_RNN}). 
	
	From Fig. \ref{fig:eva1_1_a}, we can see that ``FNN + one-step DQN'' cannot achieve optimal sum throughput within the 100 thousand simulated basic slots. ``FNN + n-step DQN'' did even worse (we leave the detailed investigation of why that is the case for the future). By contrast, ``FNN + RB-DQN'' can achieve near-optimal sum throughput.  As we can see from Fig. \ref{fig:eva1_1_b}, after replacing FNN with RNN, ``RNN + one-step DQN'' and ``RNN + RB-DQN'' can both achieve near-optimal sum throughput. Furthermore, compared with using FNN, using RNN allows faster convergence and smoother throughput with less jitters. Between ``RNN + one-step DQN'' and ``RNN + RB-DQN'', we notice that the latter has faster convergence---specifically, ``RNN + RB-DQN'' needs around 3500 basic slots to achieve the near-optimal performance, while ``RNN + one-step DQN'' needs more than 10000 basic slots to do that. 
	
	The single LSTM layer in the RNN structure, when unfolded in time, corresponds to  $ M $ layers of computation (see Fig. \ref{fig:FNN_RNN}). We next explore if  FNN can achieve performance comparable to RNN when we increase the number of hidden layers $ h $ in the FNN structure.  Fig. \ref{fig:eva1_2} compare the results between ``RNN + RB-DQN'' and ``FNN + RB-DQN''. The RNN is the same as in Fig. \ref{fig:eva1_1_a} and \ref{fig:eva1_1_b}, while the number of hidden layers of FNN varies. In particular, we set $ h=2, 11, 21, 41$. For $ h= 11, 21, 41 $,  the FNN is the residual network structure as in \cite{yu2017deep}. The reason to use the residual network sturcture is to avoid potential overfitting due to large number of hidden layers \cite{he2016deep}.
	
	As can be seen from Fig. \ref{fig:eva1_2}, ``FNN + RB-DQN'' with more hidden layers cannot achieve performance comparable to that of ``RNN + RB-DQN'' either. As mentioned earlier in Section \ref{Sec:Methodology}, the causal relationship between different elements in the input are explicitly modeled into RNN but not FNN. Perhaps this allows the RNN to search within a narrower solution space for a good solution (i.e., RNN only needs to learn within a smaller solution space, allowing it to learn a good solution in a more focused manner).

	\subsection{CS-DLMA versus WiFi}\label{Sec:eva2}
	We next compare the performances of CS-DLMA and WiFi in heterogeneous networks. As in Section \ref{Sec:eva1}, we consider the co-existence with a $ q $-ALOHA node and a TDMA node. The setups of the $ q $-ALOHA node and the TDMA node are the same as in Section \ref{Sec:eva1}. For CS-DLMA, we adopt ``RNN + RB-DQN''.  We then replace the DLMA node by a WiFi node and run the experiment again.  For the WiFi node, the carrier sensing slot and the backoff slot of WiFi are both set to one basic slot, the initial window size is set to 2, and the maximum backoff stage of WiFi is set to 2. The packet length of WiFi node varies from 1 basic slot to 4 basic slots. As a side note, we did try WiFi with different initial window sizes, maximum backoff stages, and packet lengths, but found no substantial performance difference among different settings. To conserve space, here we only present the results with varying packet lengths.
	\begin{figure}[!t]
		\subfigure[{\scriptsize Cumulative sum throughputs}]{
			\label{fig:eva2_1}
			\begin{minipage}[t]{0.24\textwidth}
				\centering
				\includegraphics[scale=0.29]{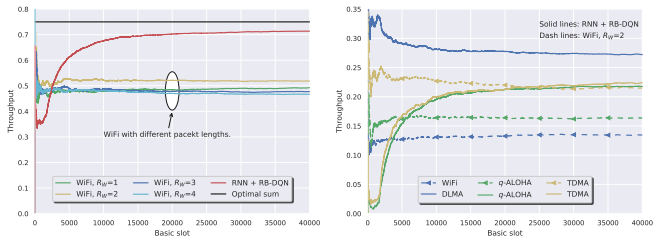}
				\vspace{-0.5in}
		\end{minipage}}
		\hspace{-0.3cm}
		\subfigure[{\scriptsize Individual cumulative throughputs}]{
			\label{fig:eva2_2}
			\begin{minipage}[t]{0.24\textwidth}
				\centering
				\includegraphics[scale=0.29]{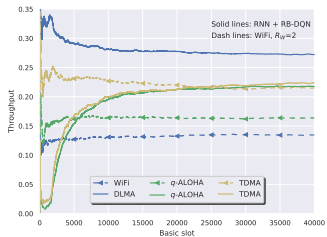}
				\vspace{-0.5in}
		\end{minipage}}
		\hspace{-0.3cm}
		\caption{\textcolor{black}{Long-term cumulative sum throughputs and individual cumulative throughputs  when one DLMA/WiFi node co-exists with  one $q$-ALOHA node and one TDMA node. The DLMA node adopts ``RNN + RB-DQN''  in both Fig. \ref{fig:eva2_1} and Fig. \ref{fig:eva2_2}. The packet length $R_W$ of the WiFi node varies from 1 to 4 in Fig. \ref{fig:eva2_1} and is fixed to 2 in Fig. \ref{fig:eva2_2}.} }
		\label{fig:eva2}
		\vspace{-0.22in}
	\end{figure}
	
	As can be seen from Fig. \ref{fig:eva2_1}, CS-DLMA can approach the near-optimal sum throughput while WiFi fails to do so.  For further details, Fig. \ref{fig:eva2_2} presents the individual throughputs of different nodes in the ``RNN + RB-DQN'' experiment and the  ``WiFi, $R_W$=2'' experiment.  As we can see from Fig. \ref{fig:eva2_2}, the individual throughputs of different nodes of ``RNN + RB-DQN'' are larger than the corresponding individual throughputs of ``WiFi, $R_W$=2''. That is, compared with the WiFi node, the DLMA node not only manages to achieve higher throughput for itself, but also to allow higher throughputs for the $ q $-ALOHA node and TDMA node. 
	
	As mentioned earlier in this paper, the carrier-sensing, collision avoidance, and backoff mechanism of WiFi are designed for a homogeneous network in which all nodes are WiFi nodes. For our case here, for example, WiFi has no mechanism to detect the repetitive channel access pattern of TDMA and to avoid the time slots occupied by the TDMA node. CS-DLMA, on the other hand, is based on an RL mechanism that has means to learn the channel access patterns of other nodes. 
	\subsection{CS-DLMA under Different Heterogeneous Network Settings}\label{Sec:eva3}
	We next investigate the performance of ``RNN + RB-DQN'' under different heterogeneous network settings. We first consider the co-existence of one DLMA node with one ALOHA node, wherein ALOHA node could adopt possibly different variants of ALOHA protocols. The ALOHA node has a packet length of 4 basic slots, i.e.,  $ R_A=4 $. 
	
	We then consider a setup in which one DLMA node co-exists with one $ q $-ALOHA node and one TDMA node. Unlike in Section \ref{Sec:eva1}, the $ q $-ALOHA node and the TDMA node now have different packet lengths with $ R_A=2 $, and   $ R_T=4 $. As in Section \ref{Sec:eva1}, the transmission probability  $ q $ of $ q $-ALOHA is fixed to 0.4 here, and the TDMA node transmits in 2 TDMA slots out of each TDMA frame of 5 TDMA slots. 
	\begin{figure*}[!t]
		\subfigure[{\scriptsize DLMA and $ q $-ALOHA}]{
			\label{fig:eva3_1_a}
			\begin{minipage}[t]{0.24\textwidth}
				\centering
				\includegraphics[scale=0.27]{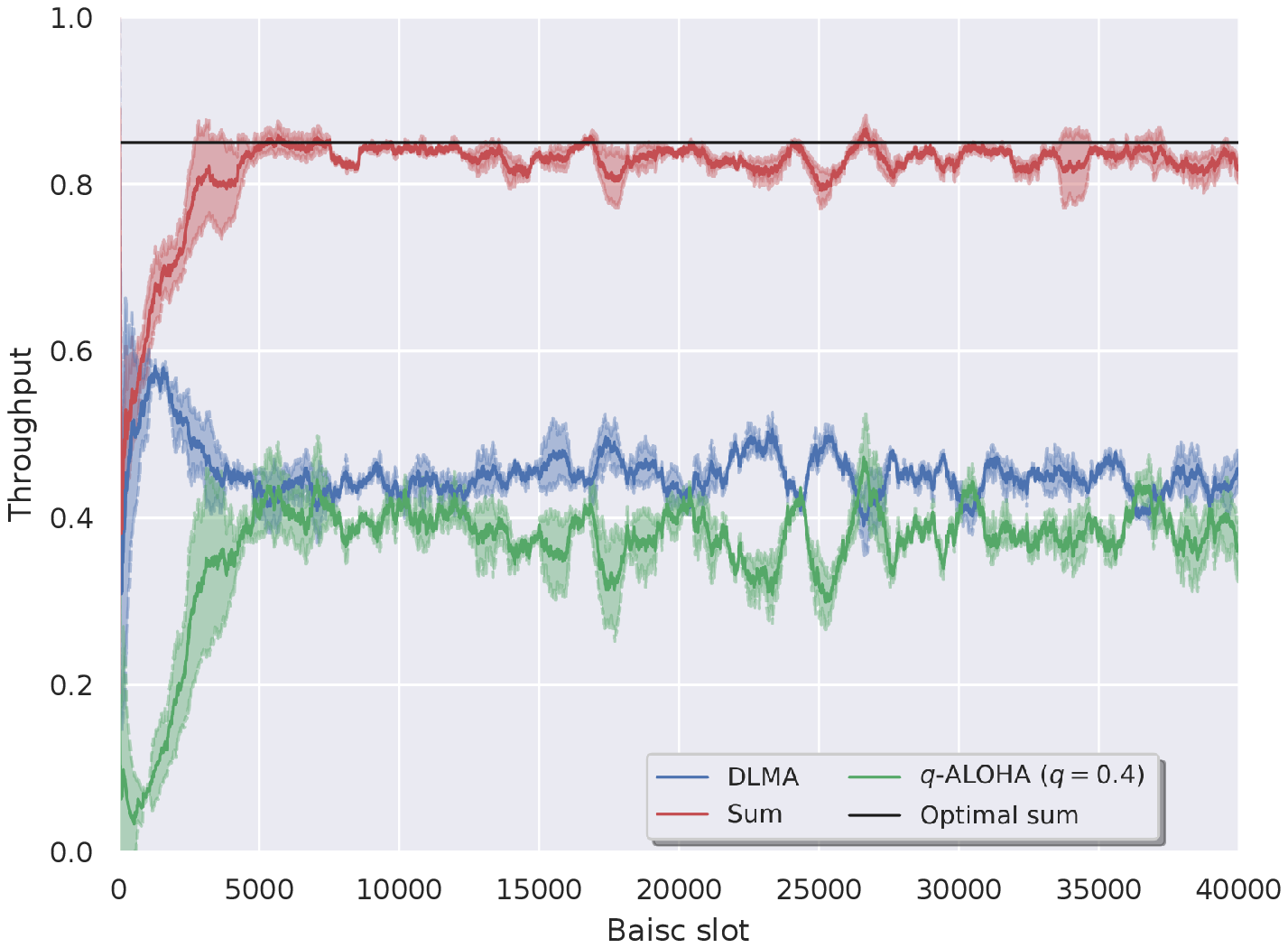}
		\end{minipage}}
		\hspace{-0.5cm}	
		\subfigure[{\scriptsize DLMA and FW-ALOHA}]{
			\label{fig:eva3_1_b}
			\begin{minipage}[t]{0.24\textwidth}
				\centering
				\includegraphics[scale=0.27]{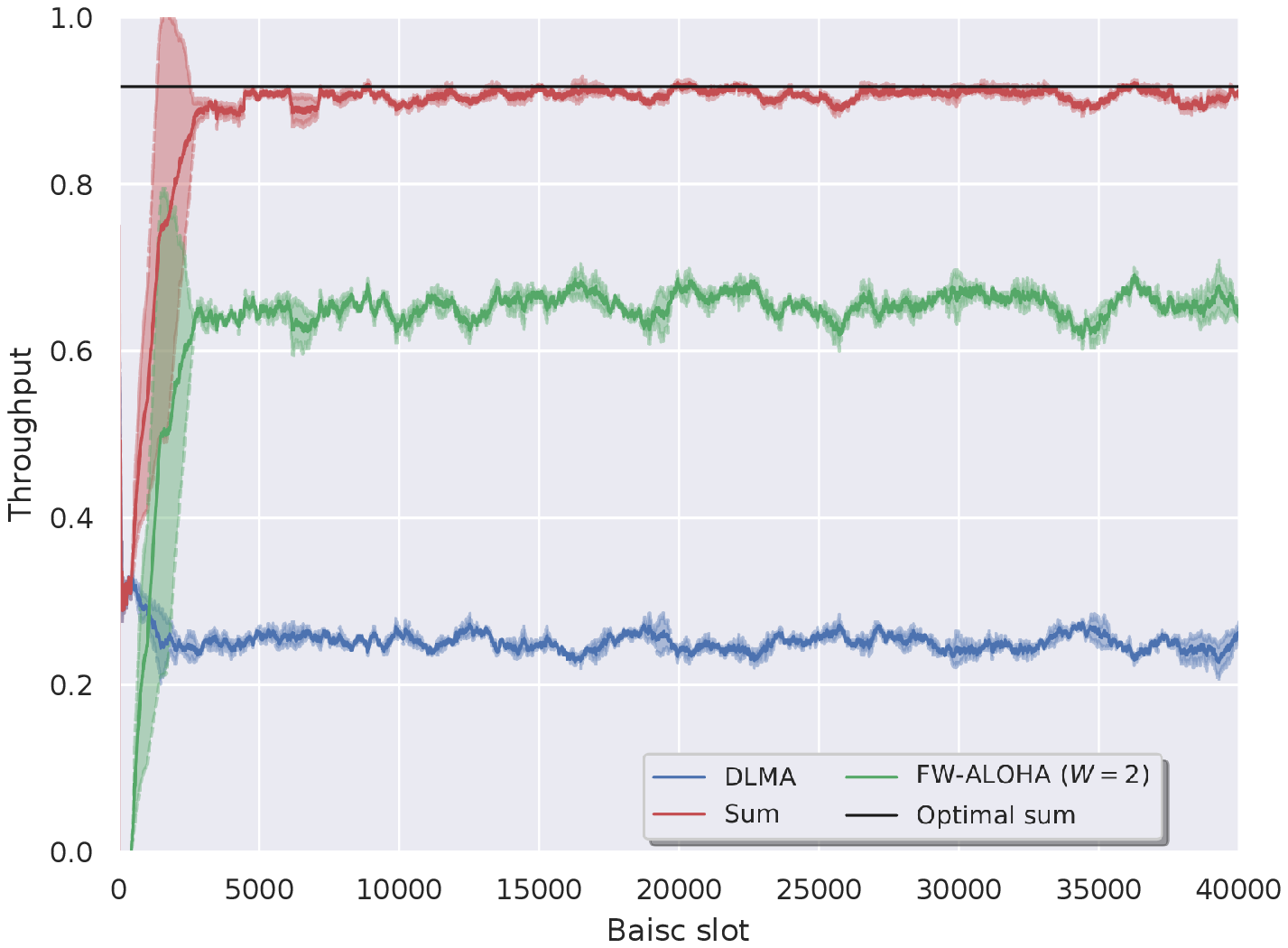}
				\vspace{-0.05in}
		\end{minipage}}
		\hspace{-0.5cm}	
		\subfigure[{\scriptsize DLMA and EB-ALOHA}]{
			\label{fig:eva3_1_c}
			\begin{minipage}[t]{0.24\textwidth}
				\centering
				\includegraphics[scale=0.27]{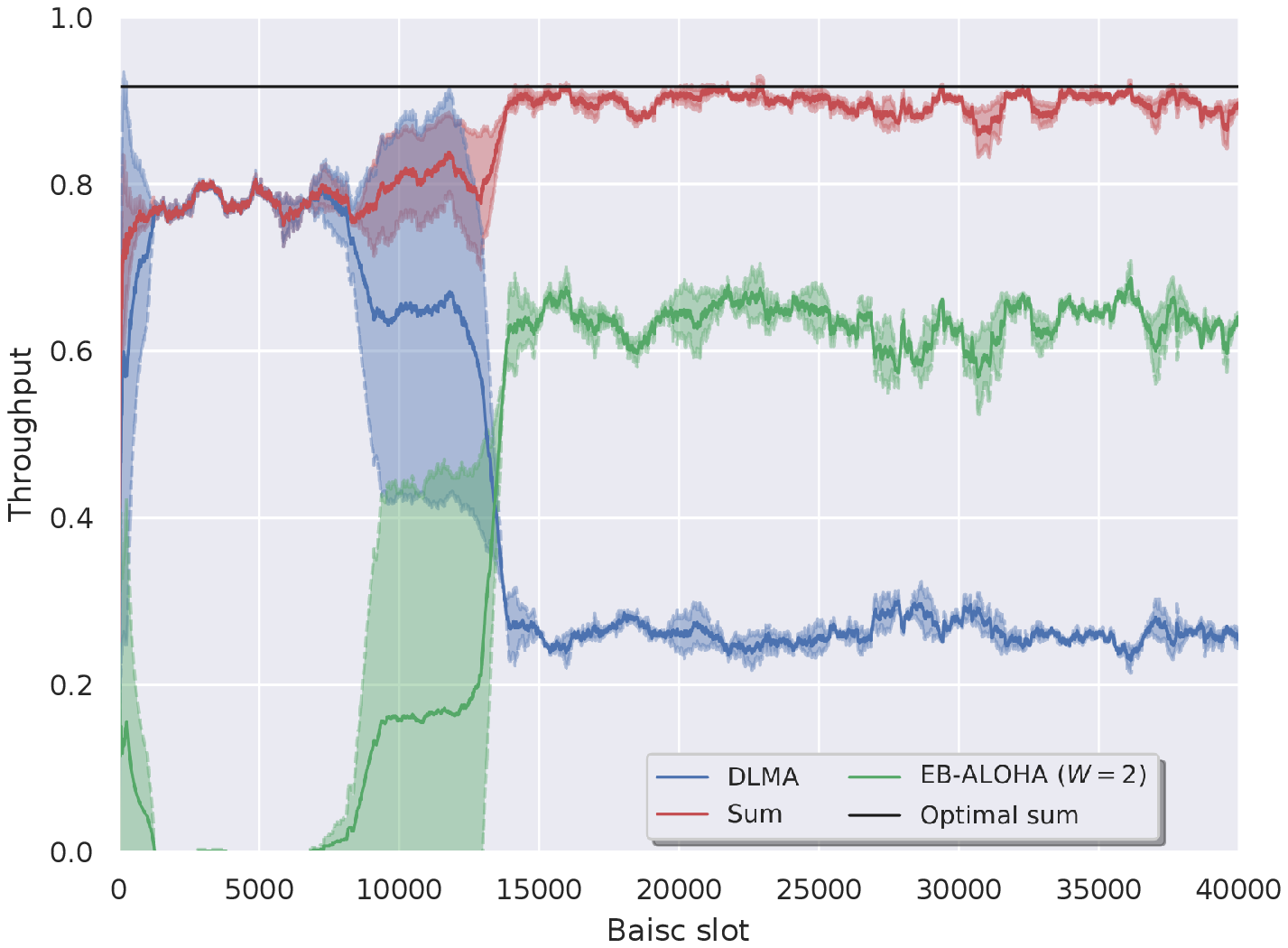}
		\end{minipage}}
		\hspace{-0.5cm}	
		\subfigure[{\scriptsize DLMA, $ q $-ALOHA and TDMA}]{
			\label{fig:eva3_2}
			\begin{minipage}[t]{0.24\textwidth}
				\centering
				\includegraphics[scale=0.27]{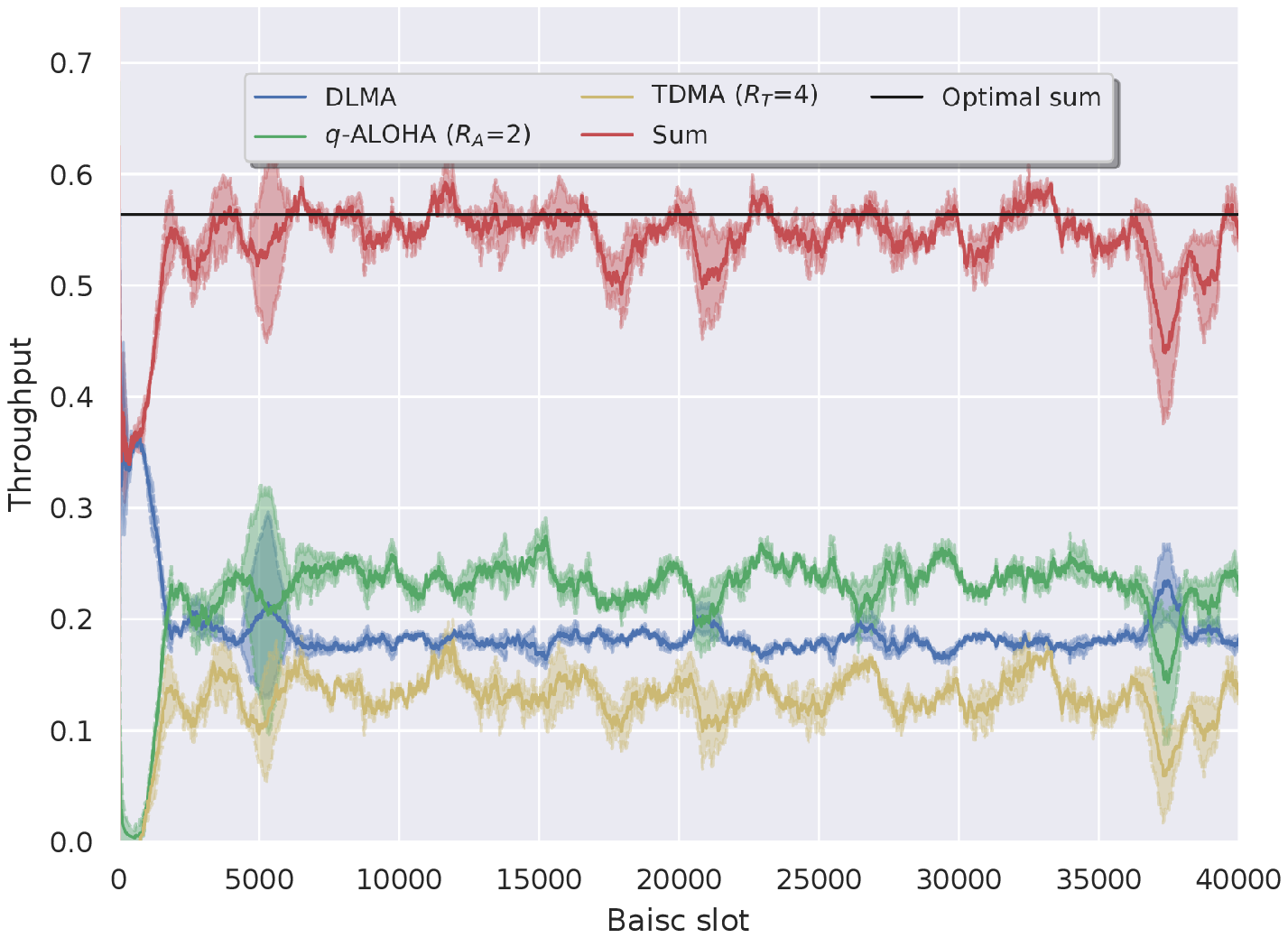}
		\end{minipage}}
		\caption{\textcolor{black}{The short-term sum throughput and individual throughputs 
				when DLMA co-exists with (a), (b), (c) different variants of ALOHA, and (d) ALOHA and TDMA. Each line (except the black line) is averaged over 4 different runs, with the shaded areas being areas within the standard deviation.}}
		\label{fig:eva3}
		\vspace{-0.21in}
	\end{figure*}
	
	\textcolor{black}{Fig. \ref{fig:eva3} presents the short-term sum throughputs and individual throughputs for the above settings. In particular, the DLMA node co-exists with one $q$-ALOHA node (with $q=0.4$) in Fig. \ref{fig:eva3_1_a}; co-exists with one FW-ALOHA node (with window size  $ W=2 $)  in Fig. \ref{fig:eva3_1_b}; co-exists with one EB-ALOHA node (with initial window size  $ W = 2$ and the maximum backoff stage $m=2$) in Fig. \ref{fig:eva3_1_c}; co-exists with one $q$-ALOHA node and one TDMA node in Fig. \ref{fig:eva3_2}. As we can see from these figures, near-optimal sum throughputs can be achieved in all cases.}
	\vspace{-0.05in}
	\section{Conclusion}
	In this paper, we showed that deep reinforcement learning (DRL) techniques  can be used to design efficient MAC protocols for heterogeneous networking. In particular, in a heterogeneous network consisting of nodes adopting different MAC protocols (e.g., ALOHA, TDMA), a node that makes use of a MAC protocol based on DRL can learn to maximize the sum throughput of all nodes in the heterogeneous environment. Furthermore, a salient feature of our DRL MAC is that it does not need to know the operating mechanisms of the co-existing MACs and the numbers of nodes using the other MACs. The DRL MAC learns to maximize the sum throughput by trial-and-error interactions with these other MACs. 
	
	We refer to our proposed DRL MAC as deep-reinforcement learning multiple access (DLMA). Compared with our past work on DLMA \cite{yu2018deep, yu2017deep}, the current work introduces carrier sensing into DLMA to further improve its efficiency and flexibility. We refer to this new class of DLMA as carrier-sense DLMA (CS-DLMA).  We demonstrated in this paper that CS-DLMA is more suitable for heterogeneous networking than WiFi MAC, a popular legacy protocol also with the carrier sensing capability. In particular, we showed that CS-DLMA can learn to co-exist with $ q $-ALOHA and TDMA to achieve near-optimal sum throughput while WiFi cannot. 
	
	This paper also investigated several variants of CS-DLMA in which different neural networks and different reinforcement learning techniques are adopted. We found that, in general, recurrent neural networks (RNN) can allow CS-DLMA to achieve higher sum throughput and faster convergence than feedforward neural networks (FNN) can. 
	
	As far as reinforcement learning is concerned, this paper focused on the techniques of deep Q-network (DQN) \cite{DQNpaper}. We studied the conventional one-step DQN and n-step DQN \cite{DQNpaper, mnih2016asynchronous}. In addition, we also put forth a new technique referred to as \textit{reward-backpropagation} DQN (RB-DQN). We showed that RB-DQN can achieve near-optimal sum throughput with faster convergence than one-step DQN and n-step DQN can. Furthermore, RB-DQN using RNN can achieve near-optimal sum throughput in different heterogeneous network settings (e.g., the co-existence of CS-DLMA with different variants of ALOHA, and the co-existence of CS-DLMA with $ q $-ALOHA and TDMA with different packet lengths). 
	
	\bibliographystyle{IEEEtran}
	\bibliography{WCNC2019}
\end{document}